\documentclass[amssymb,prl,twocolumn,superscriptaddress,floats,showkeys,amsmath]{revtex4-1}

\usepackage{epsfig}
\usepackage{bm}
\usepackage{color}
\usepackage{array}

\begin{document}

\title{Ultrafast valley-selective coherent optical manipulation with excitons \\ 
in WSe$_2$ and MoS$_2$ monolayers}

\author{A.O. Slobodeniuk}
\email{aslobodeniuk@karlov.mff.cuni.cz}
\affiliation{Department of Condensed Matter Physics, Faculty of Mathematics and Physics,
Charles University, Ke Karlovu 5, CZ-121 16 Prague, Czech Republic}

\author{P. Koutensk\'{y}}
\affiliation{Department of Chemical Physics and Optics, Faculty of Mathematics and Physics,
Charles University, Ke Karlovu 3, CZ-121 16 Prague, Czech Republic}

\author{M. Barto\v{s}}
\affiliation{Central European Institute of Technology, Brno University of Technology, Purky\v{n}ova 656/123, 612 00 Brno, Czech Republic}

\author{F. Troj\'{a}nek}
\affiliation{Department of Chemical Physics and Optics, Faculty of Mathematics and Physics,
Charles University, Ke Karlovu 3, CZ-121 16 Prague, Czech Republic}

\author{P. Mal\'{y}}
\affiliation{Department of Chemical Physics and Optics, Faculty of Mathematics and Physics,
Charles University, Ke Karlovu 3, CZ-121 16 Prague, Czech Republic}

\author{T. Novotn\'{y}}
\affiliation{Department of Condensed Matter Physics, Faculty of Mathematics and Physics,
Charles University, Ke Karlovu 5, CZ-121 16 Prague, Czech Republic}

\author{M. Koz\'{a}k}
\email{kozak@karlov.mff.cuni.cz}
\affiliation{Department of Chemical Physics and Optics, Faculty of Mathematics and Physics,
Charles University, Ke Karlovu 3, CZ-121 16 Prague, Czech Republic}


\begin{abstract}
Increasing the speed limits of conventional electronics requires innovative approaches to manipulate
other quantum properties of electrons besides their charge. An alternative approach utilizes the valley
degree of freedom in low-dimensional semiconductors. Here we demonstrate that the valley degeneracy
of exciton energies in transition metal dichalcogenide monolayers may be lifted by coherent optical
interactions on timescales corresponding to few tens of femtoseconds. The optical Stark and Bloch-Siegert effects generated by strong nonresonant circularly-polarized light induce valley-selective blue
shifts of exciton quantum levels by more than 30 meV. We show these phenomena by studying the
two most intensive exciton resonances in transiton metal dichalcogenide monolayers and compare the
results to a theoretical model, which properly includes the Coulomb interaction and exciton dispersion.
These results open the door for ultrafast valleytronics working at multiterahertz frequencies.
\end{abstract}

\maketitle

Valleytronics aims at information processing and storage by utilizing the valley degree of freedom of electrons 
instead of their charge. This new quantum number is associated with the inequivalent groups of energy degenerate 
valleys of the conduction or valence bands, which are occupied by an electron or a hole. Principles allowing to 
generate, control and read imbalanced valley populations of charge carriers have been demonstrated and studied 
in several dielectric and semiconductor materials such as diamond \cite{Isberg2013,Suntornwipat2021}, silicon 
\cite{PhysRevLett.96.236801} or AlAs \cite{Gunawan2006}, in semimetallic bismuth \cite{Zhu2012} and in 
two-dimensional materials including graphene \cite{PhysRevLett.99.236809} and transition metal dichalcogenides 
monolayers (TMDs) \cite{PhysRevLett.108.196802,Cao2012}. 

Among the other materials, two-dimensional TMDs have exceptional properties, which make them attractive for 
valleytronic applications. The two typical examples of TMDs are WSe$_2$ and MoS$_2$. These materials belong to 
direct band gap 2D semiconductors with band gap minima in $\mathrm{K}^\pm$ points (valleys) of the Brillouin 
zone \cite{Splendiani2010,PhysRevLett.105.136805,Korm_nyos_2015}. Due to the spatial and time-reversal 
symmetries of the electronic wave functions in the K$^+$ and K$^-$ points determined by the symmetry point group of 
the crystal D$_{3\text{h}}$, these materials demonstrate valley-dependent optical selection rules. Resonant light with 
right- ($\sigma^+$) or left-handed ($\sigma^-$) circular polarization induces optical transitions only in 
K$^+$ or K$^-$ valley, respectively. The optically allowed transitions lead to the generation 
of so-called intravalley bright excitons, the electron-hole pairs tightly bound by the Coulomb interaction that 
strongly interact with photons. 

The exciton states in opposite valleys  have the same energies, i.e. they are doubly degenerate by valley. 
They form a basis for a two-level system and superposition of these states 
belongs to the valley pseudospin space \cite{Jones2013,Xu2014,Ye2017}. The controllable probing and manipulation 
of such states are necessary prerequisites for valleytronic devices 
\cite{PhysRevLett.108.196802,Cao2012,Mak2012,PhysRevLett.123.096803}. One of the crucial elements of the pseudospin 
operations is the control of the energies of the two-level system, i.e. the energies of the exciton states in 
each valley. 

Lifting the energy degeneracy of the excitons in K$^+$ and K$^-$ valleys may be reached 
by Zeeman-type splitting in static 
magnetic field  
\cite{Srivastava2015,Aivazian2015,Wang2016,Smolenski2016} 
or by DC Stark effect due to electric field
\cite{PhysRevX.4.011034} applied perpendicularly to the sample plane. However, these two effects are not 
practical for several reasons: (i) the fields required to observe significant shifts are extremely high, e.g. 
magnetic field of 10\,T generates Zeeman splitting of the exciton states of only 1\,meV, (ii) the application of 
static fields allows neither ultrafast operation nor high spatial resolution, (iii) the experiments are typically 
limited to low-temperatures due to small splitting.

An alternative approach, which simultaneously solves the aforementioned problems, is to use the strong coupling of 
the excitons to light fields and to lift the energy degeneracy of the exciton states by coherent optical phenomena. 
Off-resonant circularly-polarized light waves applied to TMDs generate valley-specific blue shifts of the excitonic 
resonances via the optical Stark (OS) or Bloch-Siegert (BS) effects 
\cite{Kim2014,Sie2015,Sie2017}.
The OS and BS effects can be described in the framework of a two-level system driven by light with photon energy 
$\hbar\omega_\mathrm{pump}$ detuned from the energy of the transition $E_0$. In the case of small detuning 
$E_0-\hbar\omega_\mathrm{pump}\ll E_0$, rotating wave approximation can be applied and the OS shift dominates
\cite{Autler1955,Delone1999}. 
The magnitude of the shift of the resonance energy can be expressed as $\Delta E_\text{OS}\propto 
\mathcal{E}_\text{pump}^2/(E_0-\hbar\omega_\mathrm{pump})$, where $\mathcal{E}_\text{pump}$ is the amplitude of the electric field of the pump pulse. When the pump photon energy is small, 
$E_0-\hbar\omega_\mathrm{pump} \sim E_0$, then the contribution to the energy shift due to the Bloch-Siegert effect 
\cite{Bloch1940} $\Delta E_\text{BS}\propto \mathcal{E}_\text{pump}^2/(E_0+\hbar\omega_\mathrm{pump})$ becomes similar to OS. 
The simple two-level model, which was applied for the description of coherent optical phenomena in TMDs in the past
\cite{Kim2014,Sie2015,Sie2017}, neglects the effects of the Coulomb interaction, which brings significant corrections 
to the OS and BS effects \cite{PRB2022}. Moreover, these coherent phenomena were observed only for the lowest resonance 
in system (1sA-exciton line) and in a limited number of materials \cite{Kim2014,Sie2015,Sie2017,Cunningham2019}. 
In this Letter we report on an ultrafast valley-selective control of blue spectral shifts of 1sA as well as 1sB exciton 
resonances in WSe$_2$ and MoS$_2$ monolayers via the interaction with off-resonant circularly polarized laser pulses 
with sub-50 femtosecond durations. We present a novel theoretical approach for the description of the observed effects, 
which is based on semiconductor Bloch equations (SBE) and goes beyond the simple two-level approximation, used in 
the previous works \cite{Kim2014,Sie2015,Sie2017,LaMountain2018}. This description takes into account i) the excitonic 
(many-body) nature of the observed shifts, and ii) the Rytova-Keldysh potential, i.e. the Coulomb potential in TMD
monolayer, modified due to inhomogeneity of the system \cite{Rytova1967,Keldysh1979,PhysRevB.84.085406}.
\begin{figure*}[t]
	\centering
	\includegraphics[width=\linewidth]{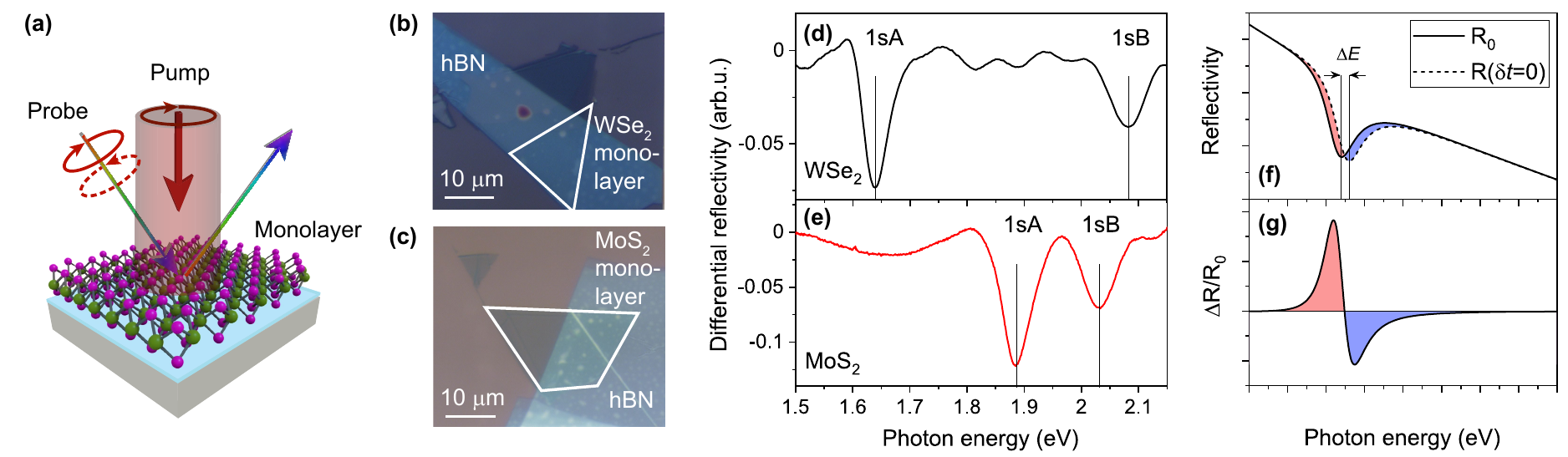}%
	\caption{(a) Layout of the experimental setup used for transient reflection spectroscopy of TMDs. The monolayer is pumped with an infrared circularly polarized pump pulse and probed by a broadband circularly polarized pulse, whose spectrum is measured. (b), (c) Optical microscope images of the studied samples of TMDs. White polygons mark the monolayers. (d), (e) Differential reflectivity spectra of WSe$_2$ (d) and MoS$_2$ (e) monolayers.  (f) 
Sketch of the monolayer reflectivities without $R_0(\hbar\omega)$ and with the pump pulse 
at zero delay time $\delta t=0$ between pump and probe pulses $R(\hbar\omega, \delta t=0)$. 
The distance $\Delta E$ between two extrema of the reflectivities manifests the shift of exciton energy due to 
pump pulse. (g) Sketch of the transient reflectance contrast
$[R(\hbar\omega,\delta t=0)-R_0(\hbar\omega)]/R_0(\hbar\omega)$. Its peak-and-dip shape explains the experimental data presented in Figs.~\ref{fig:fig_2}~c and \ref{fig:fig_3}~c.}
\label{fig:fig_1}
\end{figure*}

The studied TMDs are fabricated by exfoliation from bulk crystals. The monolayers are transferred to a Si/SiO$_2$ 
substrate and are covered by multilayer of hBN to preserve their optical properties in ambient air. In our experiments 
we measure the spectrum of transient change of reflectivity  of monolayers WSe$_2$ and MoS$_2$ as a function of the time 
delay between a femtosecond infrared pump pulse (central photon energy 0.62\,eV, FWHM pulse duration of 
$\tau_\text{pump}=38$\,fs) and 
a broadband supercontinuum probe pulse (photon energy 1.3-2.25\,eV), see Supplemental Material for the details. The layout of the experimental setup is shown in Fig.~\ref{fig:fig_1}a. 
During the experiments, the samples are imaged in situ using an optical microscope setup to ensure the spatial 
overlap of the pump and probe pulses and their position at the monolayer (each sample is about 20-30\,$\mu$m in size,
see Figs.\ref{fig:fig_1}~b and c). 
The polarization state of both pump and probe beams is controlled using broadband quarter-wave plates, which generate 
circular polarizations. The experiments are carried out at room temperature with laser repetition rate of 25\,kHz. 

Because the samples are prepared on an absorptive substrate, an increase of absorption in the monolayer corresponds to 
a decrease of reflectivity of the sample \cite{MCINTYRE1971417}. This was verified both by measuring the differential reflectivity 
$[R_0(\hbar\omega)-R_\text{sub}(\hbar\omega)]/R_\text{sub}(\hbar\omega)$  of the monolayers, 
where $R_0(\hbar\omega)$ ($R_\text{sub}(\hbar\omega)$) is the reflectivity of the sample  
(substrate), and by using finite-difference time-domain method simulations  
(for details see Ref.~\cite{PRB2022}). The energies of 1sA and 1sB exciton transitions 
obtained from our differential reflectivity measurements in the WSe$_2$ monolayer (Fig.~\ref{fig:fig_1}d) are 
$E_\mathrm{1sA}^\mathrm{WSe_2}=1.639\,\mathrm{eV}$ and
$E_\mathrm{1sB}^\mathrm{WSe_2}=2.083\,\mathrm{eV}$. In MoS$_2$ monolayers  
(Fig.~\ref{fig:fig_1}e), the excitonic transitions are shifted to $E_\mathrm{1sA}^\mathrm{MoS_2}=1.886\,\mathrm{eV}$ and 
$E_\mathrm{1sB}^\mathrm{MoS_2}=2.032\,\mathrm{eV}$. These values are 
in a good agreement with the previously measured values \cite{nano8090725} and with the theoretical calculations of the spin-orbit 
splitting energy \cite{Korm_nyos_2015}.

When the sample is illuminated by the non-resonant pump pulse, 
the excitonic transitions move to higher energies leading 
to a blue shift $\Delta E$ of the resonances in the reflectivity spectra.
The information about this shift  is encoded in reflectivity of the sample $R(\hbar\omega,\delta t)$
measured after time $\delta t$ of the pulse application (see Fig.~\ref{fig:fig_1}f). 
We consider the difference $\Delta R(\hbar\omega,\delta t)=R(\hbar\omega, \delta t)-R_0(\hbar\omega)$ 
to eliminate the contribution of the optical response of the Si/SiO$_2$ substrate.
Hence, $\Delta R(\hbar\omega,\delta t)$ contains only the contribution from the excitons in the monolayer.    
In the following we consider the experimentally measurable ratio $\Delta R(\hbar\omega,\delta t=0)/R_0(\hbar\omega)$, 
sketched in Fig.~\ref{fig:fig_1}g.
\begin{figure*}[t]
	\centering
	\includegraphics[width=\linewidth]{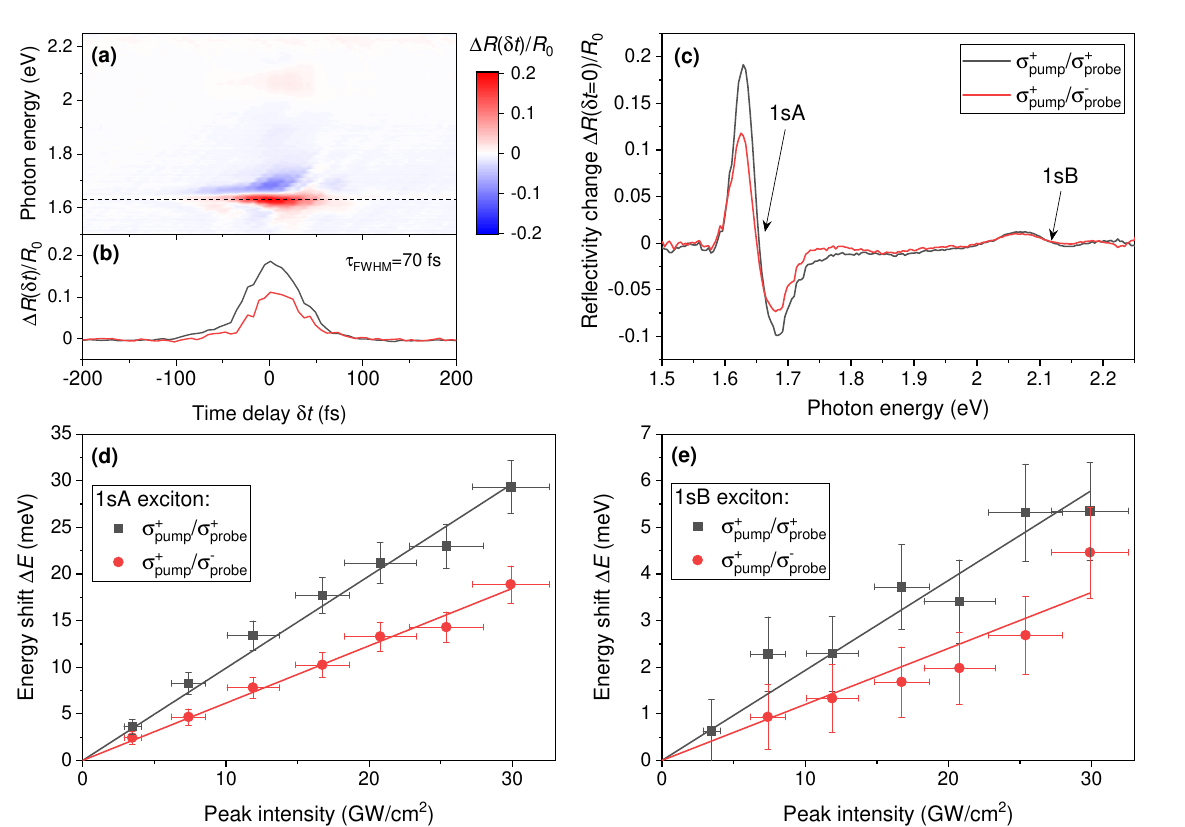}%
	\caption{(a) TR spectra of WSe$_2$ monolayer 
	$\Delta R(\hbar\omega,\delta t)/R_0(\hbar\omega)$ as a function of the time delay between pump 
	(photon energy of 0.62\,eV) and broadband probe pulses both of the same 
circular polarization handedness. (b) Profile of the signal in time-domain for the same (black curve) and opposite circular (red curve) polarization handedness of the pump and probe pulses. (c) TR spectrum at zero time delay for peak intensity of the pump pulse of 30\,$\mathrm{GW/cm^2}$. (d), (e) Maximum energy shift obtained from the measured spectra at zero time delay as a function of the peak intensity of the pump pulse for 1sA (d) and 1sB (e) excitonic resonances. }
	\label{fig:fig_2}
\end{figure*}
\begin{figure*}[t]
	\centering
	\includegraphics[width=\linewidth]{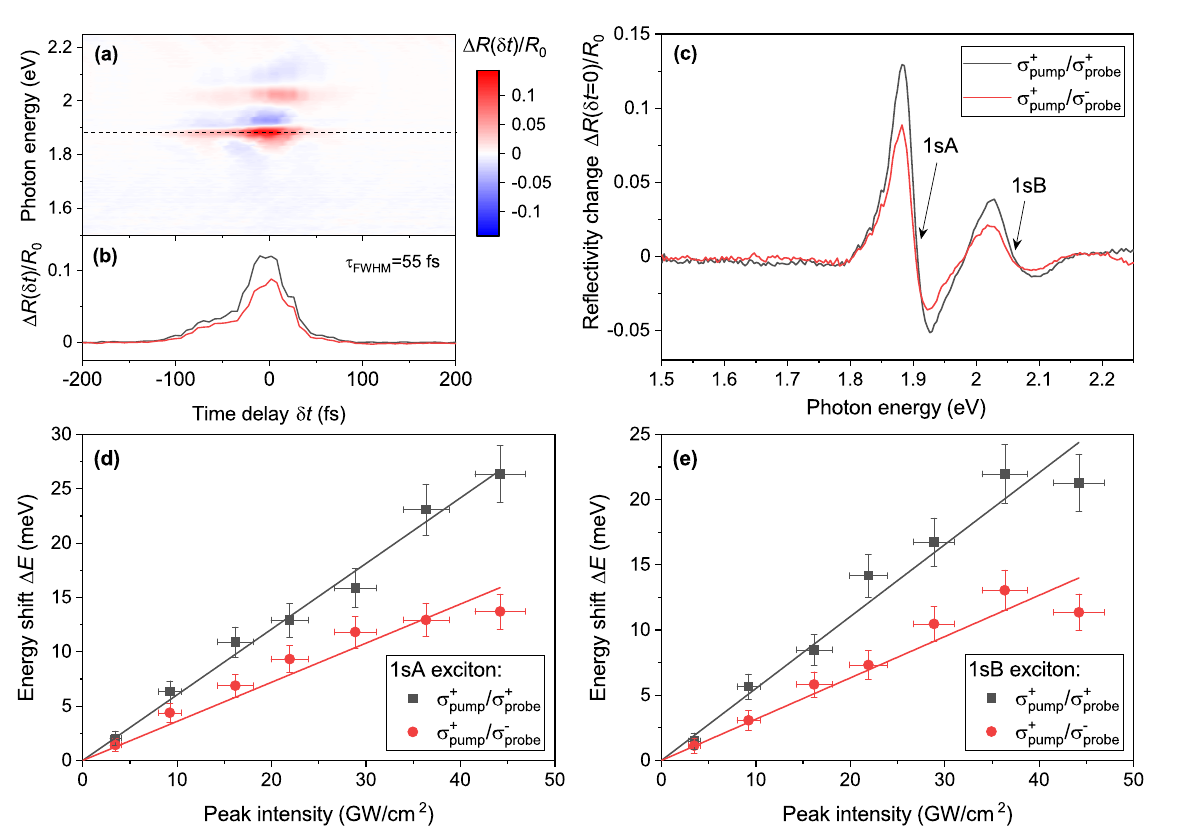}%
	\caption{(a) TR spectra of MoS$_2$ monolayer 
	$\Delta R(\hbar\omega,\delta t)/R_0(\hbar\omega)$ as a function of the time delay between the pump 
	(photon energy of 0.62\,eV) and broadband probe pulses both of the same 
circular polarization handedness. (b) Profile of the signal in time-domain for the same (black curve) and opposite circular (red curve) polarization handedness of the pump and probe pulses. (c) TR spectrum at zero time delay for peak intensity of the pump pulse of 36\,$\mathrm{GW/cm^2}$. (d), (e) Maximum energy shifts obtained from the measured spectra at zero time delay as a function of the peak intensity of the pump pulse for 1sA (d) and 1sB (e) excitonic resonances.}
\label{fig:fig_3}
\end{figure*}

In Figs.~\ref{fig:fig_2}a and \ref{fig:fig_3}a we show the delay time dependence of the transient reflectivity (TR) spectra 
$\Delta R(\hbar\omega, \delta t)/R_0(\hbar\omega)$ in monolayers WSe$_2$ and MoS$_2$. In Figs.~\ref{fig:fig_2}b and \ref{fig:fig_3}b the temporal profile of the signal is 
presented for  WSe$_2$ and MoS$_2$, respectively. 
In both materials we identify two features corresponding to the shifts of 1sA and 1sB excitons. The spectra at the zero time 
delay are plotted in Figs.~\ref{fig:fig_2}c and ~\ref{fig:fig_3}c for different combinations of circular polarization handedness of pump and probe 
pulses, namely for $\sigma_\mathrm{pump}^+/\sigma_\mathrm{probe}^+$ and 
$\sigma_\mathrm{pump}^+/\sigma_\mathrm{probe}^-$. 
The measurements with opposite combinations of the polarizations, i.e. 
$\sigma_\mathrm{pump}^-/\sigma_\mathrm{probe}^-$ and $\sigma_\mathrm{pump}^-/\sigma_\mathrm{probe}^+$, provide the same shifts in accordance with the time-reversal symmetry of TMDs \cite{PRB2022}.  

In Figs.~\ref{fig:fig_2}~d, e and \ref{fig:fig_3}~d, e we show the measured energy shifts of the exciton lines for the two combinations of circular 
polarization handedness of the pump and probe pulses as functions of the peak intensity of the pump pulse. In all cases, the 
observed blue shift changes linearly with the pump intensity. From linear fits of the data we obtained the ratios 
between the shifts caused by OS and BS effects, which are summarized in Table~\ref{tab:tab_1} and compared to the theoretical 
values. The observed deviation of the experimental and theoretical results can be explained by 
an inaccuracy of the parameters of the monolayer, used for the theoretical estimation as well as 
2D crystal's imperfections which reduce its valley-dependent optical response.     
	\begin{table}[h]
 \begin{center}
 \begin{tabular}{ccccc}
 \hline\hline \\ [-0.5ex]
 {$\chi$} & 1sA, ex. & 1sA, th. & 1sB, ex. & 1sB, th.  \\ [1.0ex]
 \hline \\
 WSe$_2$  & $1.62\pm 0.08$  & $2.48$ & $1.58\pm 0.17$ & $1.95$  \\ [1.2ex]
 MoS$_2$  & $1.67\pm 0.09$  & $2.19$ & $1.75\pm 0.14$ & $2$ \\ [1.2ex]
 \hline
\end{tabular}
\end{center}
\caption{The ratio $\chi=\Delta E_\text{OS}/\Delta E_\text{BS}$ of OS to BS shifts for 1sA and 1sB excitons, derived experimentally (ex.) and calculated theoretically (th.).}
		\label{tab:tab_1}
\end{table}

The maximum observed energy shift of 1sA exciton in both materials is about 30\,meV, which is a much larger value 
compared to the previously published results. Such a large transient shift is reached due to high peak intensity of the 
pump pulse of 30-50\,$\mathrm{GW/cm^2}$. Thanks to the short pump pulse duration ($\tau_\text{pump}$=38\,fs), the transient signal is not 
accompanied by a long signal component corresponding to the real carrier population generated by nonlinear absorption 
of the pump pulse. The maximum relative transient change of the sample reflectivity of about 20\% is promising for 
applications of this effect in valleytronic devices working on femtosecond time scales. Note that this reflectivity change is reached at room temperature and without an enhancement by an optical cavity, which could further increase the transient signal \cite{LaMountain2021}.

The theoretical description of the observed blue shifts of the exciton resonances 
is based on a perturbative solution of the SBE. Since the 1sA and 1sB exciton transitions couple 
the valence and conduction bands with the same spin, i.e. the fixed pair of the bands, 
we restrict our consideration to an effective two band model 
\cite{Korm_nyos_2015}. 
The interaction between $\sigma^\pm$
polarized light, characterized by electric field 
$\mathbf{E_\pm}=\mathcal{E}(\cos(\omega t),\pm\sin(\omega t))$, with carries 
in $\mathrm{K}^\pm$($\tau=\pm$) valleys of TMDs is defined by the Hamiltonian
$H^\tau=H^\tau_0+H^\tau_\text{int}$. Here
\begin{align}
H_0^\tau=&\sum_\mathbf{k}(E_{e,k}\alpha^{\tau\dag}_\mathbf{k}\alpha^\tau_\mathbf{k}+
E_{h,k}\beta^{\tau\dag}_{-\mathbf{k}}\beta^\tau_{-\mathbf{k}}) \nonumber \\ +& 
\sum_{\mathbf{k},\mathbf{k}',\mathbf{q}\neq0} \frac{V_\mathbf{q}}{2} (\alpha^{\tau\dag}_{\mathbf{k}+\mathbf{q}}
\alpha^{\tau\dag}_{\mathbf{k}'-\mathbf{q}}\alpha^\tau_{\mathbf{k}'}\alpha^\tau_\mathbf{k}+
\beta^{\tau\dag}_{\mathbf{k}+\mathbf{q}}\beta^{\tau\dag}_{\mathbf{k}'-\mathbf{q}}
\beta^\tau_{\mathbf{k}'}\beta^\tau_\mathbf{k})\nonumber \\-&
\sum_{\mathbf{k},\mathbf{k}',\mathbf{q}\neq0} V_\mathbf{q}
\alpha^{\tau\dag}_{\mathbf{k}+\mathbf{q}}
\beta^{\tau\dag}_{\mathbf{k}'-\mathbf{q}}\beta^\tau_{\mathbf{k}'}\alpha^\tau_\mathbf{k}
\end{align}
is the two-band Hamiltonian with included Coulomb interaction.  
The first term defines a spectrum of electrons $E_{e,k}=\hbar^2k^2/2m_e+E_\mathrm{g}$ 
and holes $E_{h,k}=\hbar^2k^2/2m_h$ in TMDs, where $k=|\mathbf{k}|$. 
$m_e,m_h>0$ are the electron and hole effective masses, 
$E_\mathrm{g}$ is the bandgap in the system, $\alpha^\tau_\mathbf{k}$ and 
$\beta^\tau_\mathbf{k}$ are the annihilation 
operators for electrons and holes, with momentum $\mathbf{k}$ in $\tau$ valley.
The remaining terms describe the Coulomb interaction in the system. 
Here $V_\mathbf{q}$ is the Fourier transform of the Rytova-Keldysh potential. 
The light-matter interaction term reads
\begin{equation}
H^\tau_\text{int}=-\mathbf{P}^\tau\,\cdot\mathbf{E_\pm}=-\sum_\mathbf{k} 
d^\tau_\mathrm{cv}\mathcal{E}^\tau_\pm(t)\alpha^{\tau\dag}_\mathbf{k}\beta^{\tau\dag}_{-\mathbf{k}} + \text{h.c.}
\end{equation}			
where $\mathbf{P}^\tau$ is the polarization operator of the system in $\tau$ valley, 
$d_\mathrm{cv}^\tau=\tau d_\mathrm{cv}$ is the transition dipole moment between the valence and conduction bands
and $\mathcal{E}^\tau_\pm(t)=\mathcal{E}\exp(\mp i\tau\omega t)$. 
Note that $H^\tau_\text{int}$ has a similar form as the light-matter interaction for a two-level system in the rotating-wave approximation. However, $H^\tau_\text{int}$ is exact and its form originates from the specific structure of the interband transition dipole moments in $\mathrm{K}^\pm$ points of TMDs.  
To describe the energy shift of the exciton transitions we introduce the
quantum average of polarization 
$P^\tau_\mathbf{k}(t)\equiv\langle\beta^\tau_{-\mathbf{k}}\alpha^\tau_\mathbf{k}\rangle$ and  
electron and hole population  $n^\tau_\mathbf{k}(t)\equiv\langle\alpha_\mathbf{k}^{\tau\dag}\alpha^\tau_\mathbf{k}\rangle=
\langle\beta^{\tau\dag}_{-\mathbf{k}}\beta^\tau_{-\mathbf{k}}\rangle$. 
For the latter equality we omit the damping and collision terms in monolayer \cite{PRB2022}.   
 The SBE read
\begin{align}
\label{eq:bloch_p}
i\frac{\partial P^\tau_\mathbf{k}}{\partial t}=&e^\tau_kP^\tau_\mathbf{k}+
(2n^\tau_\mathbf{k}-1)\omega^\tau_{R,\mathbf{k}},\\
\label{eq:bloch_n}
\frac{\partial n^\tau_\mathbf{k}}{\partial t}=&2\text{Im}\big[
\omega^{\tau*}_{R,\mathbf{k}}P^\tau_\mathbf{k}\big], 
\end{align} 
where we introduced the parameter 
$\hbar e^\tau_k(t)=E_\mathrm{g}+\hbar^2k^2/2m-
\sum_\mathbf{q} V_{\mathbf{k}-\mathbf{q}}n^\tau_\mathbf{q}$, with the exciton reduced mass $m=m_em_h/(m_e+m_h)$,
and Rabi energy $\hbar\omega^\tau_{R,\mathbf{k}}(t)=d^\tau_\mathrm{cv}\mathcal{E}^\tau_\pm(t)+
\sum_{\mathbf{q}\neq \mathbf{k}}V_{\mathbf{k}-\mathbf{q}}P^\tau_\mathbf{q}$.
We solve this set of equations for the electric field given by a 
superposition of the pump and probe fields. Namely, 
we substitute $\mathcal{E}^\tau_\pm(t)\rightarrow \mathcal{E}_\mathrm{pump}e^{-i\tau\omega_\mathrm{pump}t}+
\mathcal{E}_\mathrm{probe}e^{\mp i\tau\omega_\mathrm{probe}t}$ in $\omega_{R,\mathbf{k}}$ for 
$\sigma_\mathrm{pump}^+/\sigma_\mathrm{probe}^\pm$ geometry of the experiment.
The solutions of Eqs.~(\ref{eq:bloch_p}) and (\ref{eq:bloch_n}) are obtained by assuming 
that the contribution $\delta P^\tau_\mathbf{k}$ of the probe field  
to the polarization $P^\tau_\mathbf{k}$, generated by the pump field is small, since
$\mathcal{E}_\mathrm{probe}\ll\mathcal{E}_\mathrm{pump}$. 
Introducing $P^\tau_\mathbf{k}+\delta P^\tau_\mathbf{k}$ 
into Eq.~(\ref{eq:bloch_p}) 
we obtain the equation for $\delta P^\tau_\mathbf{k}$ in the presence of $P^\tau_\mathbf{k}$, 
which is supposed to be known (see details in \cite{PRB2022}). 
The solution of the equation at the 
1s exciton energy $E_{ex}$ has a resonant structure 
$\delta P^\tau_\mathbf{k}\propto 1/(E_{ex}+\Delta E_\tau-\hbar\omega_\textrm{probe})$, which
defines the energy shift of 1s exciton resonance 
\begin{align}
\label{eq:shift}
\Delta E_\pm=\frac{2|d_\mathrm{cv}|^2\mathcal{E}_\mathrm{pump}^2}{(E_{ex}\mp\hbar\omega_\mathrm{pump})}
\Big[\rho_{1s}+\frac{\eta_{1s}}{(E_{ex}\mp\hbar\omega_\mathrm{pump})}\Big].
\end{align}
Here $\Delta E_\pm$ correspond to the OS and BS shifts in $\mathrm{K}^\pm$ points, respectively. 
The first term of Eq.~(\ref{eq:shift}) corresponds to the {\it exciton-pump-field interaction}, 
while the second term provides a correction due to {\it exciton-exciton interaction} in the system \cite{Ell1989}. 
To evaluate $\rho_{1s}$ and $\eta_{1s}$ we use the hydrogen-like 1s wavefunction, 
which is a remarkably good approximation for the ground-state excitons \cite{PhysRevLett.123.136801}.
In this case $\rho_{1s}=16/7$, while the exciton-exciton correction is material dependent and 
reaches $\sim 10\%-20\%$ of $\rho_{1s}$ for the studied monolayers. Hence, our predicted energy shift is
approximately 2-3 times larger than the two-level approximation result \cite{Sie2017}. 
It demonstrates the importance of Coulomb interaction and many-body effects in the evaluation of the OS and BS shifts.

Using the presented theory we numerically calculated the expected energy shifts for the parameters used in our 
experiments \cite{PRB2022}. The results in form of coefficient $\kappa\equiv \Delta E/\mathcal{E}_\text{pump}^2$ are presented in Tab.~\ref{tab:tab_2}. 	
Almost all experimental results are close to the theoretical estimations.
The observed deviations of both results, in particular for 1sB excitons in WSe$_2$, 
can be explained either by an inaccuracy of the parameters of the monolayer, used for the theoretical study
or by the peculiarities of 2D crystals used in the experiment. 
\begin{table}[h]
 \begin{center}
 \begin{tabular}{ccccc}
 \hline\hline \\ [-0.5ex]
 {$\kappa\,[\mathrm{eV}\!\cdot\!\mbox{\AA}^2/\text{V}^2]$} & OS,1sA & BS,1sA & OS,1sB & BS,1sB  \\ [1.0ex]
 \hline \\
 WSe$_2$, th.  & $33.6$ & $13.5$ & $12.7$ & $6.5$ \\ [1.2ex]
 WSe$_2$, ex. & $26.4\pm 0.7$ & $16.3\pm 0.5$ & $4.1\pm 0.4$ & $2.6\pm 0.4$ \\ [1.2ex]
 MoS$_2$, th.  & $17.7$ & $8.1$  & $13.1$ & $6.4$ \\ [1.2ex]
 MoS$_2$. ex. & $16.0\pm 0.6$ & $9.6\pm 0.4$  & $14.7\pm 0.5$ & $8.4\pm 0.4$ \\ [1.2ex]
 \hline
\end{tabular}
\end{center}
\caption{The coefficient $\kappa\equiv\Delta E/\mathcal{E}_\text{pump}^2$ calculated theoretically (th.) and 
    experimentally (ex.) for OS ($\Delta E_\text{OS}$) and BS ($\Delta E_\text{BS}$) shifts of 1sA and 1sB excitons, 
		in WS$_2$ and MoS$_2$ samples.}
		\label{tab:tab_2}
\end{table}	
		
The observed valley-specific energy shifts of excitonic resonances in 2D TMDs WSe$_2$ and MoS$_2$ allow to lift the valley degeneracy in these materials at extremely short time scales of several tens of femtoseconds. The observed maximum relative transient change of the reflectivity of the WSe$_2$ monolayer at the 1sA exciton resonance reaches 
20\%. Together with the large circular dichroism of 
$\xi_\text{max}=(\Delta R_{\sigma^+/\sigma^+}-\Delta R_{\sigma^+/\sigma^-})/\Delta R_{\sigma^+/\sigma^+}=38\%$ 
($\Delta R_{\sigma^+/\sigma^+}$ and $\Delta R_{\sigma^+/\sigma^-}$ are reflectivity changes of the probe pulse for the co- and counter-rotating circularly polarized pump and probe fields) and the nonresonant ultrafast operation without a significant population of real excitons, this effect is promising for ultrafast valleytronic applications. Furthermore, by applying circularly polarized pump pulses with low photon energy and high intensity, strong-field nonlinear coherent phenomena may lead to generation of valley-polarized currents observable in transport \cite{Kundu2016,JinemezGalan2020}
or optical \cite{Langer2018} measurements.

\section{Acknowledgments}

The authors would like to acknowledge the support by Czech Science
Foundation (project GA23-06369S) and Charles University (UNCE/SCI/010, SVV-2020-260590,
PRIMUS/19/SCI/05). M. Barto\v{s} acknowledges the support by the ESF under the project
CZ.02.2.69/0.0/0.0/20 079/0017436.


%

\end{document}